\documentclass[aps,prl,twocolumn,preprintnumbers,amsmath,amssymb]{revtex4}

\usepackage{graphicx}
\usepackage[colorlinks=true,citecolor=blue]{hyperref}

\newcommand{\la}{\langle}
\newcommand{\ra}{\rangle}

\newcommand{\mcal}{\mathcal}
\newcommand{\f}{\frac}

\newcommand{\te}{\text}
\newcommand{\tb}{\textbf}
\newcommand{\omg}{\omega}

\newcommand{\Omg}{\Omega}
\newcommand{\sig}{\sigma}

\newcommand{\var}{\varepsilon}

\newcommand{\ua}{\uparrow}
\newcommand{\da}{\downarrow}
\newcommand{\Ua}{\Uparrow}

\newcommand{\ga}{\gamma}
\newcommand{\Ga}{\Gamma}

\begin{document}
    %\linenumbers
    \title{Fast, High Fidelity Quantum Dot Spin Initialization without a Strong Magnetic Field by Two-Photon Processes}
    \author{Arka Majumdar$^1$}
    \email{arkam@stanford.edu}
    \author{Ziliang Lin$^1$}
    \author{Andrei Faraon}
    \author{Jelena Vu\v{c}kovi\'{c}}
    \affiliation{E.L.Ginzton Laboratory, Stanford University, Stanford, CA, 94305\\ $^1$ Equal
    Contribution}
    %\date{\today}
\begin{abstract}
We describe a proposal for fast electron spin initialization in a
negatively charged quantum dot coupled to a microcavity without
the need for a strong magnetic field. We employ two-photon
excitation to access trion states that are spin forbidden by
one-photon excitation. Our simulation shows a maximum
initialization speed of $1.3$ GHz and maximum fidelity of $99.7\%$
with realistic system parameters.
\end{abstract}

\maketitle

Recent demonstrations of cavity quantum electrodynamics (QED)
effects with semiconductor quantum dots (QDs) coupled to
microcavities %\cite{article:eng07}-\cite{article:hen07}
\cite{article:eng07,article:hen07,article:sri07,article:press07,article:yoshio04}
show that these systems are robust and scalable platforms for
quantum information science. The QD-cavity QED experiments
performed so far treat each QD as a two-level quantum system
consisting of a ground state and the single exciton excited state.
However, multilevel quantum systems are necessary for fast,
complex, and high fidelity quantum information processing
\cite{article:ima99}. Each of these systems should contain two
stable ground states between which the transition is dipole
forbidden and one or more excited states that serve as a passage
for population transfer between the ground states ($\Lambda$
system) \cite{article:press08,article:gup08}. Initialization,
coherent manipulation, and readout of these $\Lambda$ systems are
essential for quantum information processing including controlled
phase gate \cite{article:dua04,article:fus08}, quantum repeaters
and networks \cite{article:cir97}, and remote entanglement
distribution or creation \cite{article:cho05}. Therefore,
realizing a multilevel system in a QD coupled to a cavity is
crucial for semiconductor cavity QED implementation of quantum
computers.

One way to realize a $\Lambda$ system in a negatively charged QD
is by employing the Zeeman-split electron spin states as the
ground states and a trion state as the excited state
\cite{article:ima99}. Recent experiments with QDs (not coupled to
a cavity) have demonstrated spin initialization with magnetic
field along the QD growth direction (Faraday geometry)
\cite{article:ata06} as well as with field perpendicular to the QD
growth direction (Voigt geometry)
\cite{article:xu07,article:ema07,article:press08}.
%\cite{article:xu07}-\cite{article:press08}.
Initialization in the Faraday geometry requires weak magnetic
field and can achieve very high fidelity, but the initialization
is slow because the process relies on random spin-flip Raman
scattering. On the other hand, initialization in the Voigt
geometry is fast, but a strong magnetic field is required to mix
the spin states and split the resulting eigenstates for high
fidelity. The high cost and large space required to achieve a
strong magnetic field present a significant drawback of this
initialization method.

In this Letter, we propose a fast and high fidelity spin
initialization method for a negatively charged QD coupled to a
microcavity. Our method also works for a positively charged QD
\cite{article:ger08}, but we will focus on the negatively charged
one as an example in this paper. The major advantage of our method
is the absence of a strong magnetic field, because there is no
need to mix the electron or the hole spin states. Instead, we
employ two-photon process to excite transitions that are otherwise
spin forbidden by one-photon excitation. A doubly resonant cavity
is used to enhance both the excitation and the spontaneous
emission rates. We show that our method can achieve a maximum
initialization speed of $1.3$ GHz and fidelity of $99.7\%$ with
realistic system parameters.

\begin{figure}
    \includegraphics[scale=.6]{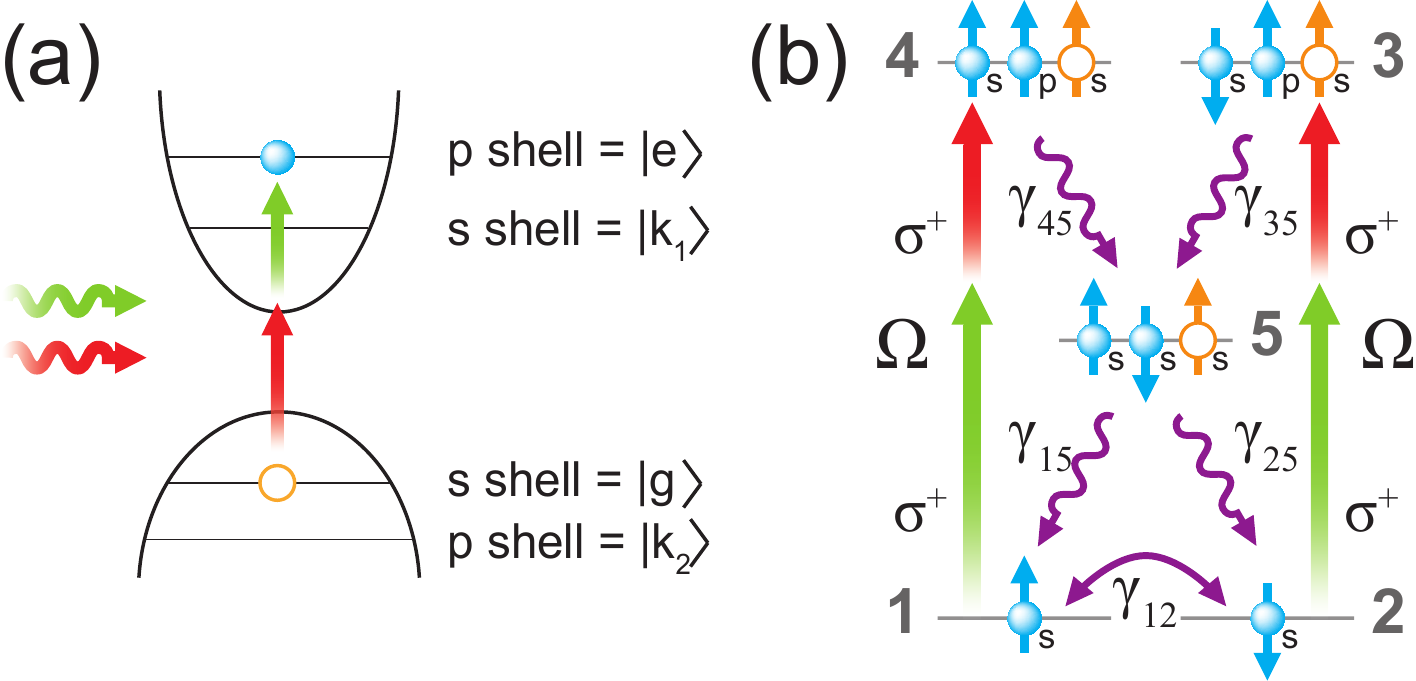}
\caption{(color online) (a) Quantized energy levels of a neutral
QD. The potential in the $xy$ plane (perpendicular to the QD
growth axis) is modeled as a two-dimensional harmonic oscillator.
Upon two-photon absorption, an electron is excited from the
valence band \textit{s} orbital ($|g\rangle$) to the conduction
band \textit{p} orbital ($|e\rangle$). (b) Energy level diagram of
relevant electron and trion states in a negatively charged QD.
Electrons (holes) are shown in solid (opened) circles. Subscripts
indicate orbital labels. The ground levels are
$|\ua_s\ra\equiv|1\ra$ and $|\da_s\ra\equiv|2\ra$. The first
excited state is the trion state $|\ua_s\da_s\Ua_s\ra\equiv|5\ra$.
The second excited states are trion states excited by two-photon
processes $|\ua_s\ua_p\Ua_s\ra\equiv|4\ra$ and
$|\da_s\ua_p\Ua_s\ra\equiv|3\ra$. The straight arrows indicate
two-photon excitation, the curly arrows indicate various decays of
excited states, and the curved arrow indicates spin
flip.}\label{fig:method}
\end{figure}

We consider a strongly confined QD and neglect the mixing between
the light hole and heavy hole bands because of their large energy
splitting. We model the QD potential as a finite well along its
growth axis ($z$ axis) and a two-dimensional oscillator
perpendicular to its growth axis ($xy$ plane)
\cite{article:kor09,article:rei08}. The effective mass for
electron (hole) is $m_e^*$ ($m_h^*$) and the oscillator frequency
for the conduction (valence) band is $\omg_e$ ($\omg_h$). The
harmonic oscillator quantized levels are labeled as \textit{s}
orbital, \textit{p} orbital, etc. (Fig.~\ref{fig:method}a).
Because the QD is much more confined along the $z$ axis than in
the $xy$ plane, the first several excited states have
wavefunctions occupying different states in the $xy$ plane but the
same ground state along $z$ axis \cite{book:mic03}.

Fig.~\ref{fig:method}b shows the relevant QD energy levels for our
initialization method. The ground levels are the electron spin up
$|\ua_s\ra$ and spin down $|\da_s\ra$ in the \textit{s} orbital.
The application of a weak magnetic field of $0.2$ T reduces
hyperfine induced spin-flip rate $\gamma_{12}$ but does not mix
the two spin states \cite{article:elz04,article:kro04}. The first
excited state is the trion state $|\ua_s\da_s\Ua_s\ra$ with the
electrons in the conduction band \textit{s} orbital and the hole
in the valence band \textit{s} orbital. The second excited states
are trion states generated by two-photon excitation
$|\ua_s\ua_p\Ua_s\ra$ and $|\da_s\ua_p\Ua_s\ra$ in which one of
the electrons is in the conduction band \textit{p} orbital.

The spin initialization to state $|\ua_s\ra$ is performed as
follows: Two-photon excitation with $\sig^+$ circularly polarized
light transfers population from $|\ua_s\ra$ to
$|\ua_s\ua_p\Ua_s\ra$ and from $|\da_s\ra$ to
$|\da_s\ua_p\Ua_s\ra$ with a rate $\Omg$. The electron in the
conduction band \textit{p} orbital then relaxes to the conduction
band \textit{s} orbital by a phonon assisted process with a
typical timescale of $10$-$30$ ps \cite{book:mic03}. Due to Pauli
exclusion principle, the resulting state is $|\ua_s\da_s\Ua_s\ra$.
The population in this state will preferentially decay to the
state $|\ua_s\ra$ with rate $\Gamma\equiv\gamma_{15}$ because the
decay to state $|\da_s\ra$ is spin forbidden. Eventually, the
population accumulates in state $|\ua_s\ra$. Two-photon excitation
cannot directly create an electron and a hole in the conduction
and valence band \textit{s}-orbitals as these two orbitals have
different parity. For this reason, we use \textit {p}-orbital for
intermediate states.

In order to achieve fast initialization, a doubly resonant cavity
is needed to enhance the QD spontaneous decay rate $\Gamma$ and
the two-photon excitation rate $\Omg$. We consider InAs/GaAs QDs
with \textit{s}-\textit{s} transition wavelength of $930$ nm.
Since the \textit{p} orbital is approximately $50$ meV above the
\textit{s} orbital in the conduction band \cite{book:mic03}, this
implies that the \textit{s}-\textit{p} transition wavelength is
approximately $900$ nm. Therefore, we need a doubly resonant
cavity at $930$ nm and $1800$ nm to enhance the spontaneous decay
rate and the two-photon excitation rate respectively. Both
resonant modes should be polarization degenerate, because all the
selection rules described in this Letter require circularly
polarized light.

We follow Ref.~\cite{article:lin09} to estimate the two-photon
excitation rate of a QD inside a microcavity. When both photons
are of the same frequency, the cavity-enhanced two-photon
absorption rate is $\Gamma_{\omg_c}^\te{TPA} =
2\pi|\Omg|^2\delta(\omg_d-2\omg)$, where the effective Rabi rate
$\Omega$ is
%\begin{linenomath}
\begin{equation*}
    \Omg=\f{\eta PQ\phi}{4\hbar^2\omg n^2\var_0V}
    \sum_k\f{|\tb d_{gk}||\tb d_{ke}|\psi_{gk}\psi_{ke}}{\Delta_{gk}},
\end{equation*}
%\end{linenomath}
where $\eta$, $P$, $Q$, $n$, and $V$ are the light coupling
efficiency into the cavity, excitation power (measured outside the
cavity), cavity quality factor, refractive index of material that
the cavity is composed of, and cavity mode volume, respectively.
$\omg_c$, $\omg_d$, and $\omg$ are the cavity resonance frequency,
the transition frequency between two QD levels coupled by the
two-photon process, and the laser frequency, respectively. $\tb
d_{gk}$ ($\tb d_{ke}$) is the dipole moment between the ground
(excited) state and the intermediate state $k$.
$\Delta_{gk}=\omg_k-\omg_g-\omg$ is the laser detuning from the
transition between the QD ground state and the intermediate state.
The delta function $\delta(\omg_d-2\omg)$ represents energy
conservation. The factor $\phi$ describes the spectral mismatch
between the laser and the cavity, $\phi(\omega) =
\f{\omg/\omg_c}{1+4Q^2(\omg/\omg_c-1)^2}$. The factor $\psi_{gk}$
($\psi_{ke}$) characterizes the reduction in Rabi rate due to
position mismatch between the QD and the electric field maximum
within the cavity, $\psi_{gk} = \f{|\tb E(\tb r)|}{|\tb E(\tb
r_M)|}\left(\f{\tb d_{gk}\cdot\hat{\tb e}}{|\tb d_{gk}|}\right)$,
where $\tb E(\tb r)$, $\tb E(\tb r_M)$, and $\hat{\tb e}$ are the
electric field at $\tb r$, electric field maximum, and electric
field polarization at the QD location, respectively.

For the transition from valence band \textit{s} orbital to
conduction band \textit{p} orbital, the dominant intermediate
states are the conduction band \textit{s} orbital $|k_1\ra =
|s_c\ra$ and the valence band \textit{p} orbital $|k_2\ra =
|p_v\ra$ (Fig. \ref{fig:method}a). In our analysis, we assume
$|\tb d_{gk_1}|=|\tb d_{ek_2}|=e|\tb r_{cv}|$ ($|\tb r_{cv}|=0.6$
nm \cite{article:eli00}), which is the transition moment between
the valence band and the conduction band integrated over a unit
cell. We also calculate $|\tb d_{ek_1}|=el_e$ and $|\tb
d_{gk_2}|=el_h$, where $e$ is the electronic charge and $l_e$
($l_h$) is the oscillating length of the electron (hole) with
$l_e=\sqrt{\hbar/(2m_e^*\omg_e)}$
($l_h=\sqrt{\hbar/(2m_h^*\omg_h)}$). With $\eta=2\%$, $Q=5000$,
$V=(\lambda/n)^3$, $\phi=1$, $\psi=1$, and $P=50\ \mu$W, we
calculate $\Omg/2\pi=4.5$ GHz. We note that we use very
conservative number of $2\%$ for in-coupling efficiency into the
cavity, as in our previous experiments on controlling cavity
reflectivity with a single QD \cite{article:eng07}, but
significantly higher coupling efficiency exceeding $80\%$ is
possible with fiber tapers or cavity-waveguide couplers
\cite{article:Andrei07,article:fiber05}. Therefore, much higher
$\Omega$ is possible with a smaller excitation power.

The fidelity and speed of our spin initialization method are
analyzed using method based on the master equation
\cite{book:gar05}. Here we analyze the case when $\sigma^+$ light
is applied. The time evolution of the density matrix $\rho$ of the
QD is given by
%\begin{linenomath}
\begin{equation*}
    \f{d\rho}{dt}=-\f i\hbar[\mcal H,\rho]+\sum_j\mcal{L}_j(\rho),
\end{equation*}
%\end{linenomath}
with the Hamiltonian
%\begin{linenomath}
\begin{equation*}
    \mcal H =
    \hbar\Omg(|\ua_s\ua_p\Ua_s\ra\la\ua_s|+|\da_s\ua_p\Ua_s\ra\la\da_s|)+\te{h.c.}.
%    \mcal H=
%    \begin{pmatrix}
%    0 & 0 & 0 & \Omg & 0 \\
%    0 & 0 & \Omg & 0 & 0 \\
%    0 & \Omg & 0 & 0 & 0 \\
%    \Omg & 0 & 0 & 0 & 0 \\
%    0 & 0 & 0 & 0 & 0
%    \end{pmatrix}
\end{equation*}
%\end{linenomath}
The Zeeman splitting between the states $|\ua_s\ra$ and
$|\da_s\ra$ is approximately $5.4\ \mu$eV for an electronic
$g$-factor of $0.46$ \cite{article:ema07} and a magnetic field of
$0.2$ T. This splitting is much smaller than $\hbar\Omg$ and
$\hbar\Gamma$, therefore we neglect it in our analysis. $\mcal
L_j(\rho)\equiv D_j \rho D_j^\dag - \f12D_j^\dag D_j\rho-\f12\rho
D_j^\dag D_j$, where $D$'s are the collapse operators:
%\begin{linenomath}
\begin{align*}
    D_1 &= \sqrt\Ga|\ua_s\ra\la\ua_s\da_s\Ua_s|, \\
    D_2 &= \sqrt{\ga_{25}}|\da_s\ra\la\ua_s\da_s\Ua_s|, \\
    D_3 &= \sqrt{\ga_{35}}|\ua_s\da_s\Ua_s\ra\la\da_s\ua_p\Ua_s|, \te{ and}\\
    D_4 &= \sqrt{\ga_{45}}|\ua_s\da_s\Ua_s\ra\la\ua_s\ua_p\Ua_s|.
\end{align*}
%\end{linenomath}
Numerical values of the rates used for the simulation are
$\ga_{25}/2\pi=100$ kHz \cite{article:ata06}, $\ga_{12}/2\pi=10$
kHz, and $\ga_{35}/2\pi=\ga_{45}/2\pi=8$ GHz corresponding to a
relaxation time of $20$ ps (Fig.~\ref{fig:method}b). The QD
spontaneous emission rate $\Gamma$ is Purcell enhanced and its
numerical value ranges from $5$ to $20$ GHz, corresponding to a
Purcell factor ranging from $50$ to $200$.

The fidelity and initialization speed have strong dependence on
$\Omg$ and $\Gamma$. Experimentally, $\Omg$ is tuned by changing
the excitation light intensity and by employing cavity resonance
at two-photon excitation frequency; $\Gamma$ is tuned by placing
the QD in nanocavities of various quality factors. To find the
fidelity, the steady-state density matrix is solved. The fidelity
is defined as the population in state $|\ua_s\ra$, namely
$\rho_{11}$. To find the initialization speed, the density
operator is evolved in time from the initial condition of
$\rho_{11}(t=0) = \rho_{22}(t=0)=1/2$, where $\rho_{22}$ is the
population in the state $|\da_s\ra$. The initialization time is
defined as the time required for $\rho_{11}(t)$ to reach the value
of $1-1/e$.
\begin{figure}
    \includegraphics[scale=.3]{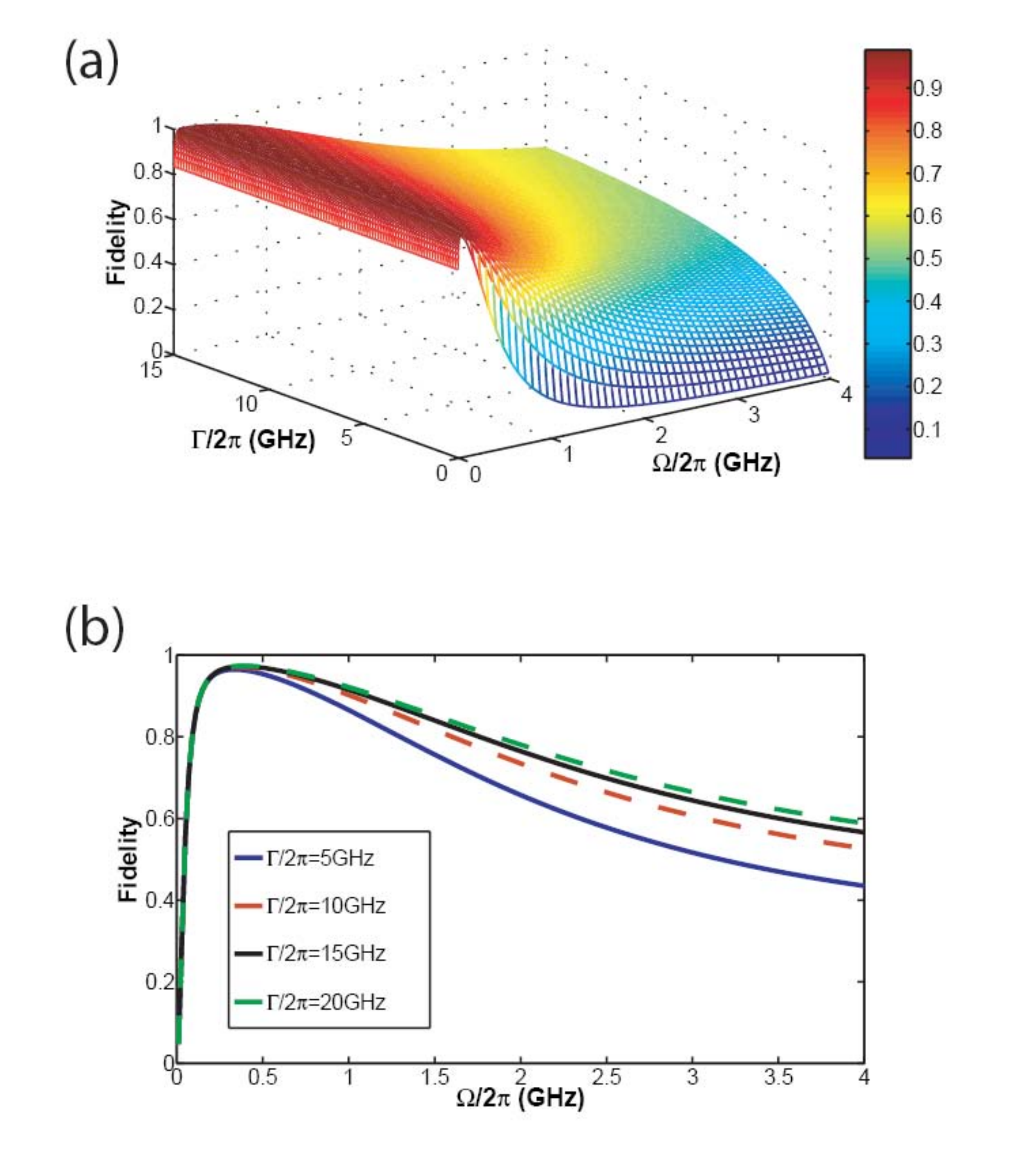}
\caption{(color online) (a) Fidelity of electron spin
initialization as a function of  two-photon excitation rate
$\Omega/2\pi$ and spontaneous emission rate $\Gamma/2\pi$. (b)
Slices through plot (a) for specific $\Gamma/2\pi$ values,
indicating maximum attainable fidelity of $99.7\%$.}
\label{fig:fidelity}
\end{figure}

Fig.~\ref{fig:fidelity}a shows the initialization fidelity as a
function of the two-photon effective Rabi rate $\Omg$ and the
one-photon spontaneous emission rate $\Gamma$. For fixed $\Gamma$,
increasing $\Omg$ depletes population from state $|\da_s\ra$ more
efficiently and therefore leads to increasing population in state
$|\ua_s\ra$ and thus higher fidelity. However, while depleting
population from state $|\da_s\ra$, two-photon excitation also
depletes population from state $|\ua_s\ra$
(Fig.~\ref{fig:method}b). Therefore, there are two competing
processes: one depletes the population in $|\uparrow_s\rangle$
with a rate of $\Omega$ and the other populates
$|\uparrow_s\rangle$ with a rate of $\Gamma$. Hence, increasing
$\Omega$ indefinitely, for a fixed $\Gamma$ depletes
$|\uparrow_s\rangle$ and ultimately decreases the fidelity of
initialization (Fig.~\ref{fig:fidelity}a,b). Such reduction in
fidelity is more prominent for lower $\Gamma$. On the other hand,
for fixed $\Omg$, increasing $\Gamma$ allows faster population
transfer from state $|\ua_s\da_s\Ua_s\ra$ to state $|\ua_s\ra$,
leading to higher fidelity. The maximum fidelity obtained is
$99.7\%$. This fidelity is comparable to the fidelity of
initialization achieved in the Faraday geometry $(\geq 99.8\%)$
\cite{article:ata06} at a magnetic field of $0.2$ T and in the
Voigt geometry $(\approx 99.7\%)$ at a magnetic field of $1$ T
\cite{article:ema07}. It should be noted that the spin flip rate
$\gamma_{12}$ also limits the fidelity. For a spin flip rate
increased from $\gamma_{12}/2\pi=10$ kHz (as in Fig.
\ref{fig:fidelity}) to $100$ kHz and $1$ MHz, the maximum fidelity
decreases to $99\%$, and $98\%$ respectively.

\begin{figure}
    \includegraphics[scale=0.3]{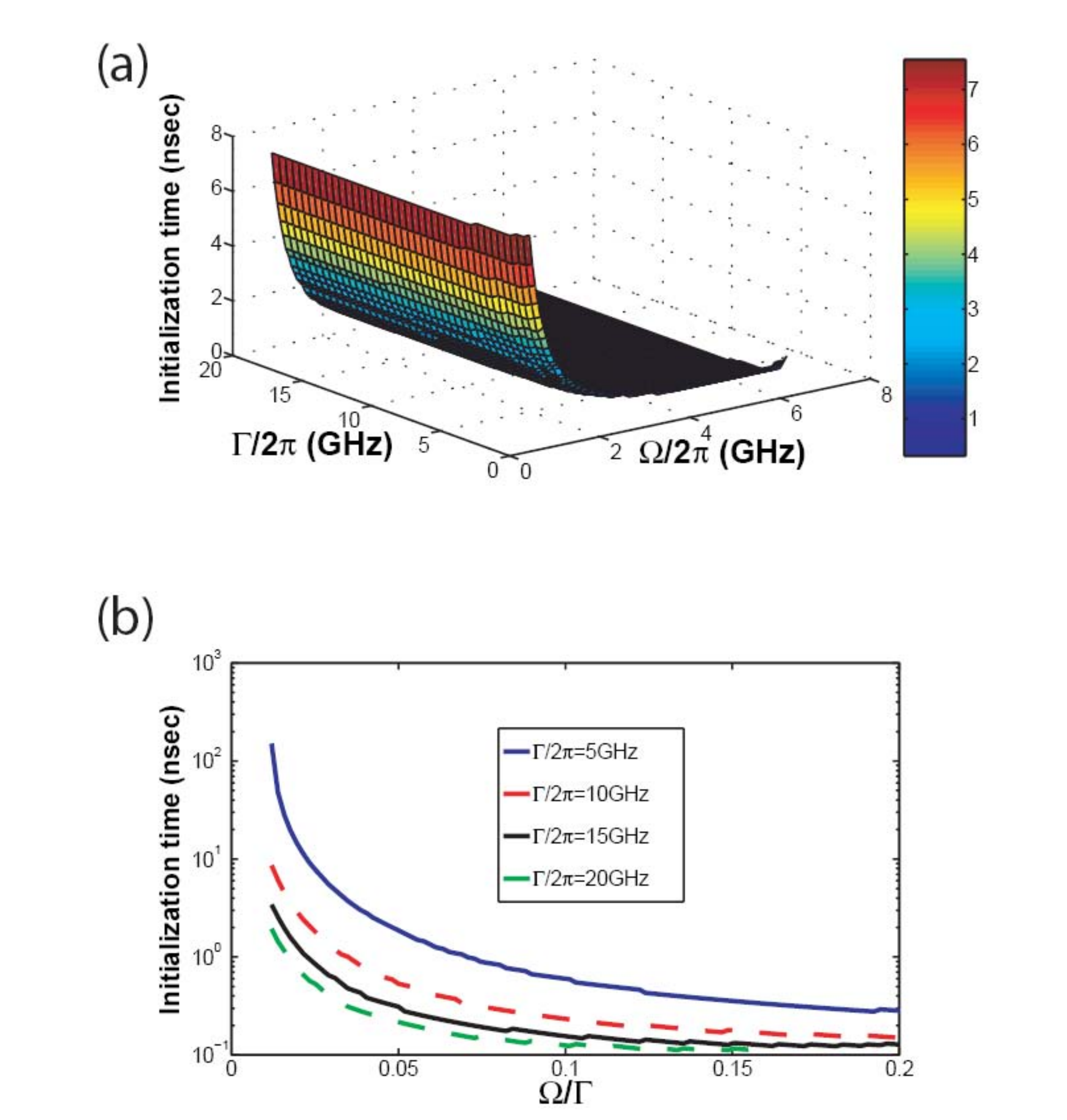}
    \caption{(color online) (a) Time needed to perform electron spin initialization as a function of $\Omega$ and $\Gamma$.
(b) Initialization time as a function of the ratio of
    $\Omega$ and $\Gamma$ for different values of $\Gamma$.}\label{fig:speed}
\end{figure}

Fig.~\ref{fig:speed}a shows the time needed to perform spin
initialization as a function of $\Omega$ and $\Gamma$. As shown in
Fig.~\ref{fig:fidelity}b, an increase in $\Omega/\Gamma$ results
in decrease in the fidelity . In fact, for $\Omega/\Gamma
> 0.2$, the fidelity never reaches the value of $1-1/e$ and
therefore initialization time is arbitrarily long. For this
reason, we study only the range $\Omega/\Gamma<0.2$ in
Fig.~\ref{fig:speed}b. It should be emphasized that the value
$0.2$ is obtained for the parameters used in the simulation, and a
different value will be obtained for a different set of
parameters. As shown in Fig.~\ref{fig:speed}b, when
$\Omega/\Gamma<0.2$, the initialization speed increases with
increasing $\Omega/\Gamma$. Although increasing $\Gamma$ in
general increases the speed, this speed increase is not
significant when $\Gamma\gg\ga_{35}=\ga_{45}$ because the phonon
assisted relaxation rates from $|\ua_s\ua_p\Ua_s\ra$ and
$|\da_s\ua_p\Ua_s\ra$ to $|\ua_s\da_s\Ua_s\ra$ become the limiting
factor. The minimum initialization time calculated is $120$ ps,
giving the maximum speed of $1.3$ GHz. This speed of
initialization is much higher than the maximum speed obtained in
the Faraday geometry ($100$ kHz) \cite{article:ata06} and the
Voigt geometry ($144$ MHz) \cite{article:ema07}. It should be
noted that the speed in Voigt geometry can potentially be enhanced
by using a cavity, but that is not true for Faraday geometry.
Unlike fidelity, a change in spin-flip rate $\gamma_{12}$ does not
affect the speed of initialization, as
$\gamma_{12}<<\Omega,\Gamma$.

It is important to note the trade off between high speed and high
fidelity. Fig.~\ref{fig:speed} shows that high two-photon
excitation rate $\Omg$ is needed to achieve high speed, but
subsequently the fidelity decreases at high $\Omg$
(Fig.~\ref{fig:fidelity}). Therefore, the speed of initialization
corresponding to the maximum fidelity of $99.7\%$ is lower
($\approx 50$ MHz). Similarly, at the highest initialization rate,
the fidelity of only $80\%$ can be achieved. However, we can still
achieve moderate fidelity with moderate speed: for example with
$\Omg/2\pi=0.5$ GHz and $\Gamma/2\pi=10$ GHz we can achieve a
fidelity of $97.3\%$ and a speed of $320$ MHz.

In conclusion, we have described a new method for initializing
electron spin in a negatively charged QD using cavity enhanced
two-photon excitation. The highest speed of $1.3$ GHz and the
maximum fidelity of $99.7\%$ can be achieved with weak magnetic
field of approximately $0.2$ T and realistic system parameters.
Although it is not possible to achieve both the maximum speed and
fidelity simultaneously, with $\Omg/2\pi=0.5$ GHz and
$\Gamma/2\pi=10$ GHz, one can achieve a fidelity of $97.3\%$
together with a speed of $320$ MHz. The use of weak magnetic field
reduces the cost and the space requirements for a quantum network,
as magnetic field of $0.2$ T can easily be achieved with a small
electromagnet, as opposed to using a large superconducting magnet
needed for strong magnetic field in other spin initialization
techniques.
\begin{acknowledgments}
The authors gratefully acknowledge financial support provided by
the Army Research Office, National Science Foundation, and Office
of Naval Research. A.M. was supported by the Stanford Graduate
Fellowship (Texas Instruments fellowship). Z. L. was supported by
the NSF Graduate Fellowship and Stanford Graduate Fellowship.
\end{acknowledgments}

\bibliography{draft4}
\end{document}